\documentclass[%
 aps,
 prb,%
 amsmath,amssymb,
 floatfix,
reprint,%
showpacs,
superscriptaddress
]{revtex4-1}

\usepackage{graphicx}
\usepackage{dcolumn}
\usepackage{bm}
\usepackage{hyperref}

\begin{document}

\title{Optimization of edge state velocity in the integer quantum Hall regime}

\author{H. Sahasrabudhe}
  \email{hsahasra@purdue.edu.}
\affiliation{Department of Physics and Astronomy, Purdue University, West Lafayette, IN}
\affiliation{Network for Computational Nanotechnology, Purdue University, West Lafayette, IN}
\author{B. Novakovic}
\affiliation{Network for Computational Nanotechnology, Purdue University, West Lafayette, IN}
\author{J. Nakamura}
\affiliation{Department of Physics and Astronomy, Purdue University, West Lafayette, IN}
\author{S. Fallahi}
\affiliation{Department of Physics and Astronomy, Purdue University, West Lafayette, IN}
\author{M. Povolotskyi}
\affiliation{Network for Computational Nanotechnology, Purdue University, West Lafayette, IN}
\author{G. Klimeck}
\affiliation{Network for Computational Nanotechnology, Purdue University, West Lafayette, IN}
\affiliation{School of Electrical and Computer Engineering, Purdue University, West Lafayette, IN}
\author{R. Rahman}
\affiliation{Network for Computational Nanotechnology, Purdue University, West Lafayette, IN}
\author{M. J. Manfra}
\affiliation{Department of Physics and Astronomy, Purdue University, West Lafayette, IN}
\affiliation{School of Electrical and Computer Engineering, Purdue University, West Lafayette, IN}
\affiliation{Station Q Purdue, Purdue University, West Lafayette, IN}
\affiliation{School of Materials Engineering, Purdue University, West Lafayette, IN}
\affiliation{Birck Nanotechnology Center, Purdue University, West Lafayette, IN}

\date{\today}

\begin{abstract}
Observation of interference in the quantum Hall regime may be hampered by a small edge state velocity due to finite phase coherence time. Therefore designing two quantum point contact (QPCs) interferometers having a high edge state velocity is desirable. Here, we present a new simulation method for realistically modeling edge states near QPCs in the integer quantum Hall effect (IQHE) regime. We calculate the filling fraction in the center of the QPC and the velocity of the edge states, and predict structures with high edge state velocity. The 3D Schr\"odinger equation is split into 1D and 2D parts. Quasi-1D Schr\"odinger and Poisson equations are solved self-consistently in the IQHE regime to obtain the potential profile near the edges, and quantum transport is used to solve for the edge state wavefunctions. The velocity of edge states is found to be \(\left< E \right> / B\), where \(\left< E \right>\) is the expectation value of the electric field for the edge state. Anisotropically etched trench gated heterostructures with double sided delta doping have the highest edge state velocity among the structures considered.
\end{abstract}

\maketitle

\section{Introduction}

Electronic interferometers have been used as tools to probe the behavior of edge states in the quantum Hall regime. A typical electronic interferometer consists of two Quantum Point Contacts (QPCs) which act as electron beam splitters, in analogy to optical interference experiments. Electrons traversing the interferometer's path accumulate an Aharonov-Bohm phase equal to $2 \pi$ times the number of magnetic flux quanta encircled. This phase can be controlled either by varying the area of the device or changing the magnetic field, yielding conductance oscillations. A major challenge for electronic interferometry is that the interfering particles must maintain phase coherence throughout their trajectory around the interference path. This is possible only if the quasiparticle edge state velocity is high enough that the time taken to traverse the interference path is smaller than the phase coherence time. Unsurprisingly, a strong correlation has been observed between the edge state velocity and the visibility of interference oscillations in the integer quantum Hall regime\cite{Gurman2016}.  

The fractional quantum Hall effect (FQHE) emerges from coulomb interactions between electrons in a two-dimensional electron gas (2DEG) in a perpendicular magnetic field \cite{Jain1989}. The FQHE states are predicted to host exotic quasiparticle excitations which carry fractional charge and obey anyonic braiding statistics, and these properties may be probed in interferometers \cite{Willett2010,Willett2013a,Ji2003,Nayak2008}. While Aharonov-Bohm interferometery has been conducted in the integer quantum Hall regime \cite{Ji2003, McClure2009, Ofek2010}, extending to the fractional quantum Hall regime has proven to be difficult.  The problem of maintaining coherence may be exacerbated in the FQH regime due to the presence of neutral edge modes, which have been predicted \cite{Sim1999, Fisher1994, Wan2002} and experimentally observed \cite{Bid2011, Inoue2014} at many states. Crucially, the neutral edge mode becomes entangled with the charge mode and must also maintain coherence along the trajectory of the interferometer\cite{Goldstein2016}, which may preclude observation of interference because the neutral modes are expected to propagate with a much lower velocity than the charge modes\cite{Lee2014, Hu2009}. Thus, optimizing device parameters to maximize the velocity of edge modes is critical to observing interference in the FQHE regime. 

\begin{figure*}
  \includegraphics{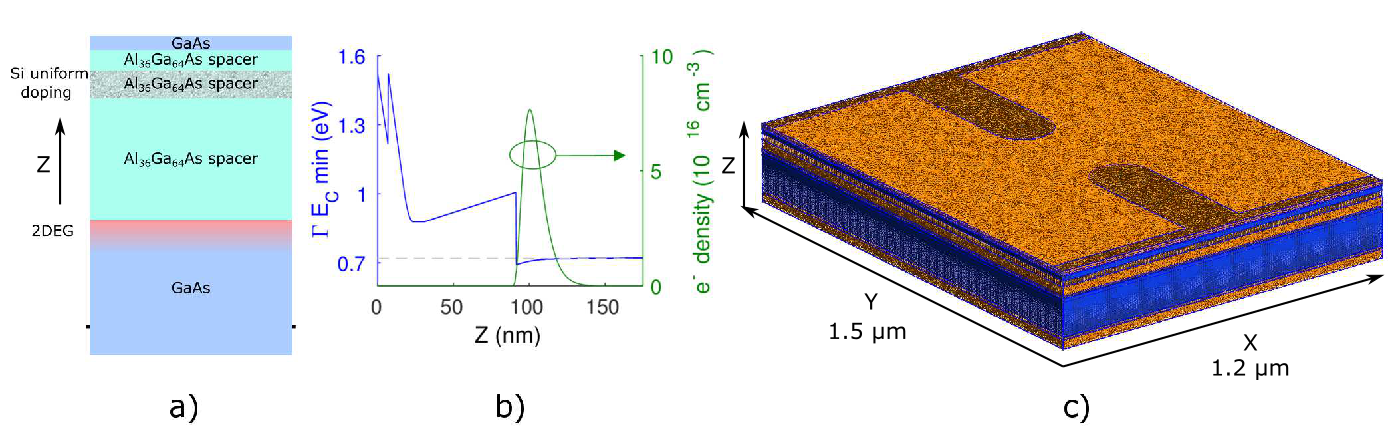}
  \caption{The details of heterostructure and gate layout used for benchmarking the simulation method. \textbf{a)} A cartoon showing the different layers and the shaded 2DEG location in the 91 nm deep single hetero-junction structure with modulation doping. A uniform doping profile with a concentration of \(4.8 \times 10^{18}\) cm\textsuperscript{-3} between 17 and 31 nm depths was used in experiments. \textbf{b)} The conduction band and electron density profiles of the quantum well solved self-consistently using a 1D Schr\"odinger-Poisson simulation. Valence band maximum on the left edge is at 0 eV, and the dashed line is the Fermi level. \textbf{c)} The \(1.5\mathrm{\mu m}\) (cross-section) \(\times\) \(1.2\mathrm{\mu m}\) (transport direction) \(\times\) \(250 \mathrm{nm}\) (growth direction) sized finite element mesh used to discretize Poisson equation for the QPC. The mesh contains tetrahedral elements (orange) to efficiently fill the dielectric regions which contain no free charge, and cuboid elements (blue) in the regions containing free charges. Pyramid and prism shaped elements (orange) are used to connect dielectric regions to charged regions. The cuboids in charged regions are thin along the growth direction, in which potential changes fast and coarse along the lateral direction in which potential changes slowly. The mesh contains \(\sim 2.16\) million points at which the electrostatic potential is solved. }
  \label{fig:structure_details}
\end{figure*}

The drift velocity of charge carriers in the classical Hall effect is equal to the ratio of electric field to magnetic field, \(E/B\). On general grounds the edge state velocity in the quantum Hall regime is expected to be proportional to the velocity scale set by \(E/B\), where in this case the electric field E is due to the confining potential at the edge. Experiments in the IQHE have confirmed that the edge velocity is approximately proportional to \(1/B\) \citep{McClure2009, Kataoka2016, Kamata2010,Kumada2011}; however, a framework for analyzing the confinement potential and predicting the velocities in different heterostructures and gating schemes is needed. While for concreteness we analyze edge velocities in the IQHE regime, the principle that edge state velocity increases for sharper confining potential is expected to generalize to the fractional regime as well \cite{Hu2009,Lee2014}. Additionally, it has been found that precise tuning of the quantum point contacts is required to achieve interference \cite{McClure2009}, so we seek to understand the behavior of QPCs both in the quantum Hall regime and at zero magnetic field. We focus on the case of GaAs/AlGaAs heterostructures with 2DEG edges defined by metallic gates. 

Numerical simulations have proven to be valuable tools for designing heterostructures\cite{Birner2006} and gated devices\cite{Arslan2008,  Siddiki2007}, as well as in explaining the results of experiments in the quantum Hall regime\cite{Panos2014,Fiori2002a}. The Poisson equation has been previously solved computationally in the IQHE regime for GaAs/AlGaAs heterostructures using a Thomas-Fermi approximation (TFA) to calculate the electron density and potential due to QPCs\cite{Eksi2010a, Siddiki2007, Arslan2008}. However in these works the doping ionization is not considered self-consistently. Also, these works model doping and quantum well regions in 2-dimensions instead of 3-dimensions. Taking these parameters into account is essential for correctly modeling the electric field at the edges, and in turn, the edge state velocity.

In this paper we present a method for calculating electron edge state velocities and electron density in gated QPCs in the IQHE regime. We use a tool which we have developed in the NEMO5\cite{Steiger2011} package for for self-consistently solving the three-dimensional Schr\"odinger and Poisson equations in the IQHE regime. Following Stopa et al.\cite{Stopa1996} and Fiori et al.\cite{Fiori2002b}, the 3D Schr\"odinger equation is split into 1D and 2D parts. The electron interactions are calculated using the mean field Hartree approximation in electrostatic simulations. We employ a frequently used incomplete ionization model for dopants in which Fermi-Dirac statistics and a donor energy level is used. We also solve the full 3D Poisson equation by accounting for the thickness of doping layers and 2DEG. Electrostatic simulations solve the potential landscape and use a Gaussian broadened Landau level density of states in the IQHE regime. The potential obtained is used in quantum transport simulations\cite{Groth2014a} to solve the 2D Schr\"odinger equation with open boundaries for the QPCs and calculate the edge state wavefunctions.

We compare the calculated conductance for QPCs with experimentally measured values to benchmark our simulations. Figure \ref{fig:structure_details} details the heterostructure and QPC gate layout used for benchmarking. On the experimental side, the heterostructure was grown by Molecular Beam Epitaxy (MBE) and Ti/Au metal gates were deposited on the surface. The devices used in the experiment have the same physical dimensions as in the simulation. We also compare the experimental 2DEG density with simulations to tune certain parameters such as donor ionization energy. The methodology used in this paper is discussed in section \ref{sec:Methodology}. The method is benchmarked with experiments in section \ref{sec:benchmark}. Sheet density, sub-band energy and edge state wavefunction profiles are discussed in the rest of section \ref{sec:results}. In section \ref{sec:optimization_of_velocity}, we study the edge state velocity as a function of magnetic field and gate voltage for four different structures in a attempt to maximize the velocity.



\section{Methodology} \label{sec:Methodology}

Typically, interference experiments operate with small source-drain biases on the order of $\mathrm{\mu V}$ to avoid heating the 2DEG  \cite{McClure2009, Camino2007}. The subband energy of electrons due to the confinement in the quantum well varies on the order of meV and is much larger than the applied bias. Source-drain bias is thus neglected in electrostatic simulations. The overall repulsive effect of electron density on itself is calculated by solving the Schr\"odinger equation self-consistently with the Poisson equation, which is a standard practice for modeling quantum systems\cite{Stern1984a}. The two equations are discretized using a non-uniform mixed element finite element mesh (Figure \ref{fig:structure_details}c). Self-consistent iterations are done using a quasi-Newton method based on the predictor-corrector method\cite{Trellakis1997}. Further details on discretization and numerics can be found in supplementary materials.  Figure \ref{fig:flowchart} shows the simulation flow.

\begin{figure}
  \includegraphics{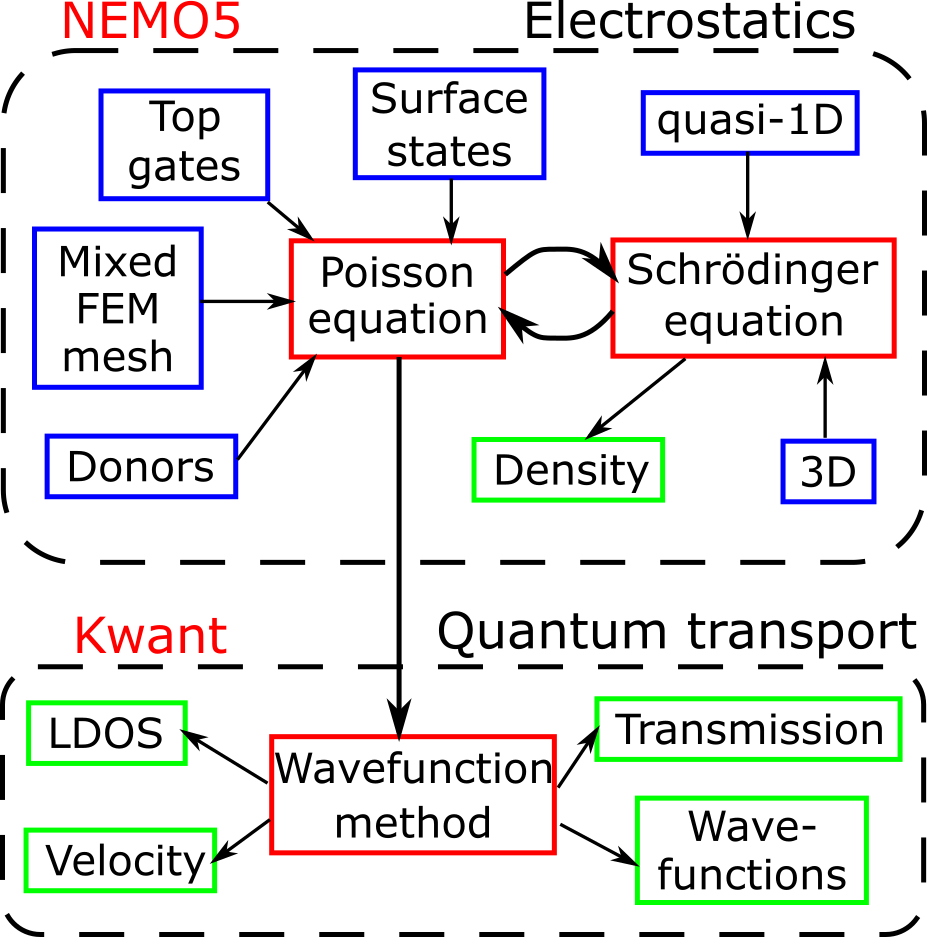}
  \caption{The simulation flowchart. Quasi-1D Schr\"odinger and Poisson equations are solved self-consistently to get the electrostatic potential and 2DEG density near the QPC. Section \ref{sec:poisson_eqn} gives details of the Poisson equation, which takes into account top gates, surface states, incomplete donor ionization and uses a mixed FEM mesh to solve complex heterostructures efficiently. The quasi-1D Schr\"odinger equation is solved for QPCs as described in section \ref{sec:schrodinger_eqn}. The potential profile obtained from electrostatics is used in quantum transport simulations (sec. \ref{sec:quantum_transport}) for calculating the QPC transmission, local density of states (LDOS), current densities, wave-functions and their velocities. Electrostatic simulations are done using the NEMO5\cite{Steiger2011} package, while the quantum transport simulations are done using the Kwant\cite{Groth2014a} package.}
  \label{fig:flowchart}
\end{figure}

\subsection{Poisson equation} \label{sec:poisson_eqn}

The effects due to the top surface, donors, gates and background disorder are included in the Poisson equation. The standard semiconductor Poisson equation with spatially varying dielectric constant is written as

\begin{equation}
\label{eq:poisson_eqn}
 - \bm{\nabla} . \left( \epsilon \bm{\nabla} \phi \right) = q (p - n + N_D^+ - N_A^-) 
\end{equation}

\noindent \(\phi\) is the electrostatic potential, \(p\) and \(n\) are hole and electron concentrations and \(N_D^+\) and \(N_A^-\) are ionized donor and acceptor concentrations respectively. \( \epsilon \) is the position dependent dielectric constant. The following expression is used to calculate the concentration of ionized donors:

\begin{equation}
\label{eq:incomplete_donor_ionization}
    N_D^+(\bm{r}) = \frac{N_D}{1+ g_D \exp \left( \frac{E_F - E_D(\bm{r})}{k T} \right)} \\
\end{equation}

\noindent \(N_D\), \(E_D\) and \(g_D\) are the total density, energy level and degeneracy respectively of donors and \(E_F\) is the Fermi level.

A single donor level is considered in the simulations and its ionization energy \( E_C - E_D \) is tuned such that the bulk 2DEG sheet density calculated from self-consistent 1D simulations matches the one obtained from experiments. The effective total density of participating Si donors \(N_D\) in eq \eqref{eq:incomplete_donor_ionization} also needs to be modified  from the real doping density used in experiments to model the effect of donor freeze out. In Al\textsubscript{x}Ga\textsubscript{1-x}As mole fractions \(\mathrm{x}>=0.36\), majority of Si atoms form deep donor levels called DX centers\cite{Schubert1984} due to the displacement of a substitutional Si atom along the [111] direction\cite{Chadi1989}. There exists a potential barrier for the trapping and de-trapping of electrons in DX centers which results in freezing out the electrons at low temperatures. Hence, the only participating donors are those which remain ionized as the device is cooled down to mK temperatures.





The Fermi level pinning on the exposed top GaAs/AlGaAs surfaces is included in the model for QPCs using Neumann boundary conditions in the Poisson equation \eqref{eq:poisson_eqn}. Charge density on the exposed surface due to occupied dangling bonds creates a Schottky barrier\cite{Wilmsen1985}. The Schottky barrier can be modeled by setting a constant voltage (usually -0.7 to -0.8 V) with respect to the Fermi level at the top surface using the Dirichlet boundary condition\cite{Iannaccone2000}. To model the Fermi level pinning, an electric field can also be specified at the top using the Neumann boundary condition\cite{Davies1995}. The electric field is calculated from the slope of the conduction band when Dirichlet boundary condition is used in 1D simulations. Setting the electric field at the top is equivalent to having frozen charges at the top surface. Other more complicated models\cite{Chen1994, Fiori2002a, Pala2002} can be used to specify a density of states at the surface, which lead to Fermi level pinning. Potential and density in the QPC channel are affected by the boundary condition of the exposed surface as shown for heterostructures with shallow 2DEG\cite{Fiori2002b}.

\subsection{Schr\"odinger equation} \label{sec:schrodinger_eqn}

The free electrons in the quantum well at the GaAs/AlGaAs interface occupy the lowest sub-band in the GaAs conduction band gamma valley and can be approximated by a parabolic dispersion relation\cite{Stern1984a}. The Schr\"odinger equation can thus be simplified using an effective-mass description:

\begin{equation}
\label{eq:schrodinger_eqn}
    - \frac{\hslash^{2}}{2} \bm{\nabla} . \left( \frac{1}{m^{*}(\bm{r})} \bm{\nabla}  \Psi(\bm{r}) \right) + V_{tot}(\bm{r}) \Psi(\bm{r})= E \Psi(\bm{r})
\end{equation}

\noindent where, \(m^{*} \left( r \right)\) is the position dependent effective-mass, which here is taken to be a constant \(0.067 m_e\) for both GaAs and AlGaAs gamma valleys. \(  V_{tot}(\bm{r}) = E_{C}(\bm{r}) - q\phi(\bm{r}) \), where \(\phi\) is the electrostatic potential in the Poisson equation \eqref{eq:poisson_eqn}. Electrostatic potential changes slower in the lateral direction than in the growth direction.  Following the work of Fiori et al\cite{Fiori2002b}, the 3D wavefunction can thus be expanded as \( \Psi \left( x,y,z \right) = \psi \left( x,y,z \right) \chi \left( x,y \right) \). Here \( \psi \) is the 1-D wavefunction along Z (growth) axis evaluated at different points in the X-Y plane. \( \chi \) is the 2D envelope along the lateral direction. The 3D Schr\"odinger equation can be separated into 1D and 2D parts by substituting the expression for \(\Psi\):

\begin{align}
\label{eq:1d_schrodinger_eqn}
  - \frac{\hbar^2}{2} \frac{\partial}{\partial z} \frac{1}{m^*} \frac{\partial \psi}{\partial z} + V_{tot} \psi &= \tilde E \psi \\
\label{eq:2d_schrodinger_eqn}
  - \frac{\hbar^2}{2 m^*} \mathbf{\nabla}_\perp^2 \chi + \tilde E_i &= E \chi
\end{align}

\noindent \( \tilde E_i \equiv \tilde E_i \left( x,y \right) \), called the sub-band energy, is the eigenvalue corresponding to \( \psi_i \). We assume the sub-band energy to be flat as it is slowly varying and calculate the approximate electron density to be used in self-consistent simulations

\begin{align}
\begin{split}
\label{eq:approx_density}
  n(x,y,z) = & \frac{k T m^*}{\pi \hbar^2} \sum_{i=1}^{\infty} \left| \psi_i \left( x, y, z \right) \right| ^2 \\
  & \times \ln \left[ 1 + \exp \left( - \frac{ \tilde E_i \left( x,y \right) - E_F}{k T} \right) \right]
\end{split}
\end{align}

\noindent The log term comes from integrating the density of states of a parabolic dispersion for 2D periodic systems.

For QPCs, equation \eqref{eq:2d_schrodinger_eqn} has open boundary conditions and needs to be solved using quantum transport algorithms (discussed in sec \ref{sec:quantum_transport}) to model the transmission, resistance, quantum current density and edge state velocity. Electrostatic simulations require the electron density for self-consistently obtaining the electrostatic potential. Using quantum transport algorithms to calculate the density in micron sized structures is computationally prohibitive. Ballistic transport simulations produce a delta function DOS in the bulk for the Landau levels, for which the energy is difficult to pinpoint, requiring a fine energy grid for integrating the electron density. Adding inelastic scattering terms to the quantum transport equations\cite{Ameen2017} for broadening the Landau level DOS would further increase the computational requirements. Making the assumption of a slowly varying sub-band energy in equation \eqref{eq:2d_schrodinger_eqn} to analytically integrate the electron density for the electrostatic simulations is computationally efficient, but neglects the lateral spread of the edge state wavefunction. The sub-band energy profile obtained from electrostatic simulations is used to solve eq. \eqref{eq:2d_schrodinger_eqn} with open boundaries. This approach has been shown to match experimental QPC conductance in zero magnetic field\cite{Fiori2002b}.

\subsubsection*{Integer Quantum Hall Regime}

The method described in the previous section is employed in the IQHE regime, where Gaussian broadened Landau level density of states is used. In the presence of a perpendicular magnetic field, equation \eqref{eq:2d_schrodinger_eqn} can be re-written as

\begin{equation}
\label{eq:2d_schrodinger_eqn_bfield}
  \frac{\left( i \hbar \bm{\nabla}_\perp + e \bm{A} \right)^2 }{2 m^*} \chi + \tilde E_i + g \mu_B B \sigma_z = E \chi
\end{equation}

\noindent \( \mathbf{A} \) is the vector potential, \(g\) is the Land\'e g-factor (discussed later), \(\mu_B\) is the Bohr magneton and B is the magnetic field. The assumption of a slowly varying \(\tilde{E}_i\) is employed again, which reduces equation \eqref{eq:2d_schrodinger_eqn_bfield} to that of non-interacting electrons trapped in 2D in presence of a perpendicular magnetic field. Then, solving \eqref{eq:2d_schrodinger_eqn_bfield} gives us the Landau level (LL) density of states (DOS) for the \(i^{th}\) sub-band

\begin{gather}
\begin{gathered}
\label{eq:landau_dos}
  D_B^i = \frac{1}{2 \pi l_B^2} \sum_{n=0}^\infty \left[ \delta \left( E - E_{i,n}^+ \right) + \delta \left( E - E_{i,n}^- \right) \right] \\
  E_{i,n}^+ = \tilde E_i + \left(n + 1/2 \right) \hbar \omega_c + g \mu_B B \\
  E_{i,n}^- = \tilde E_i + \left(n + 1/2 \right) \hbar \omega_c - g \mu_B B
\end{gathered}
\end{gather}

\noindent where, \( l_B = \sqrt{\hbar/eB} \) is the magnetic length and \( \omega_c = eB/m^* \) is the cyclotron frequency. The \( \tilde E\) term introduces a position dependence to the LL energy. This expression was used by Chklovskii et al.\cite{Chklovskii1992} for calculating electrostatics at the edges in the IQHE.

Real devices have broadened LL DOS due to disorder, collision broadening and effects due to a finite wavefunction width, which has been studied in detail by others\cite{Guven2003}. To account for these effects, a Gaussian spread DOS around the LL is used.

\begin{align}
\begin{split}
\label{eq:modified_landau_dos}
  \tilde{D}_B^i =  \frac{1}{2 \pi l_B^2} &\sum_{n=0}^\infty \frac{1}{\sqrt{2 \pi} \Delta E} \Biggl[ \exp \left( - \frac{\left(E - E_{i,n}^+ \right)^2}{2 \Delta E^2} \right) \\
  &+ \exp \left( - \frac{\left(E - E_{i,n}^- \right)^2}{2 \Delta E^2} \right) \Biggr]
\end{split}
\end{align}

\( \Delta E = \gamma \hbar \omega_c\) is a parameter than defines the spread of the states around the LL energy. The electron density can thus be written as

\begin{equation}
\label{eq:density_B_field}
  n_B \left( x,y,z \right) = \sum_{i=1}^\infty \int_{-\infty}^{\infty} \frac{\left| \psi_i \left( x, y, z \right) \right| ^2 \tilde{D}_B^i }{1+\exp \left( \frac{E_i-E_F}{kT} \right) } \, dE
\end{equation}

G\"{u}ven and Gerhardts investigated the effect of changing \( \gamma \) and temperature on the potential profile\cite{Guven2003, Gerhardts2008}. They calculated the potential profile for different values of \(\gamma\) and \(t = k T/\hbar \omega_c\).  In this paper we use the value \( \gamma = 0.05 \), which amounts to a standard deviation of \(5\%\) of the LL spacing. \(t\) is a dimensionless parameter which represents the relative strengths of thermal and magnetic energy scales. They showed that the incompressible region width  decreases by roughly \(50 \%\) when the temperature is increased from \(t=1\) to \(t=2\) for \(\gamma = 0.025\). Typical electron temperatures for interferometry experiments range from \(10-300\)mK and  magnetic fields are in the range \(0.5-10\)T\cite{Willett2009}. For these parameters, \(t\) is in the range \(5 \times 10^{-5}\) to \(3 \times 10^{-2}\). We can thus approximate the Fermi function as a step function without affecting the width of incompressible regions more than the mesh spacing. Evaluating eq. \eqref{eq:density_B_field} with a step Fermi function gives us

\begin{align}
\begin{split}
\label{eq:approx_density_B_field}
  n_B&\left( x,y,z \right) =  \frac{1}{4 \pi l_B^2} \sum_{i=1}^\infty \Biggl\{ \left| \psi_i \left( x,y,z \right) \right|^2 \\
  &\times \sum_{n=0}^\infty \Biggl[ 2 \pm \mathrm{erf} \left( \left| \frac{E_F - E_{i,n}^+}{\sqrt{2}\Delta E} \right| \right) \pm \mathrm{erf} \left( \left| \frac{E_F - E_{i,n}^-}{\sqrt{2}\Delta E} \right| \right) \Biggr] \Biggr\}
\end{split}
\end{align}

The first term in the summation comes from integrating the half Gaussian curve under the LL energy. The second term comes from integrating the density of states lying between the LL energy and the Fermi level. \(+\) or \(-\) sign is used when Fermi level is above or below  \(E_{i,n}^+\) and \(E_{i,n}^-\) respectively. Using the expression in eq. \eqref{eq:approx_density_B_field} for density helps with convergence and gradually increasing magnetic field and temperature from 0 is not required.

\subsubsection*{Land\'e g-factor}

The g-factor in bulk GaAs is 0.44; however, for two-dimensional electrons in the quantum Hall regime, spin splitting is enhanced due to exchange interactions, and experimental measurements of the spin gap in GaAs/AlGaAs heterostructures have yielded effective g-factors up to 11.65\cite{Huang2013}. The interactions between spin polarized electrons which lead to this enhanced spin splitting can be taken into account using the local spin density approximation (LSDA)\cite{Bilgec2010}. To compare the potential landscape with and without spin splitting, we performed calculations using an effective g-factor of 5.2, the same as the value used by Bilge\c{c} et al.\cite{Bilgec2010}, so that our results may be compared with the results using LSDA. We found that there is no substantial effect on the incompressible strip widths due to the low magnetic field range and Landau level broadening.

\subsection{Quantum transport} \label{sec:quantum_transport}

2D wavefunctions of the edge states can be obtained by solving equations \eqref{eq:2d_schrodinger_eqn} or \eqref{eq:2d_schrodinger_eqn_bfield} with open boundary conditions using the non-equilibrium quantum transport methods based on non-equilibrium Green's functions (NEGF)\cite{Datta1995, Lake1997} or the quantum transmitting boundary method (QTBM)\cite{Lent1990a}. This work uses the software Kwant\cite{Groth2014a} which is based on a wave-function approach similar to QTBM for solving the open system. The effective 2D Hamiltonian is discretized using finite difference method with a square grid of spacing \(a=3\) nm. The effective mass Hamiltonian with 3nm grid has a parabolic dispersion within the relevant energy range of 6 meV above the conduction band minimum in the absence of magnetic field. In the IQHE regime, the vector potential is included in the Hamiltonian using Peierls phase approximation. Landau gauge is used for calculating the vector potential since translational symmetry is required in the Hamiltonian.

Incoming and outgoing transverse modes in the QPC scattering region are calculated by solving eigenvalue problems in the cross-section away from the QPC. The 2D Schr\"odinger equation in the QPC scattering region is then solved as a linear system at a certain energy with the incoming and outgoing transverse modes as boundary conditions. Further details are present in the supplementary section and can also be referred in refs. \onlinecite{Datta1995, Groth2014a}. The wavefunction envelopes thus obtained are written as \(\chi^i_{\nu}\), where \(i\) denotes the \(i^{th}\) transverse slab and \(\nu\) is the wavefunction number. The velocity of the \(\nu^{th}\) transverse mode is then calculated using a dot product of the \(\nu^{th}\) wavefunctions \(\chi^{i \dagger}_{\nu}\) in slab \(i\) and \(\chi^{i+1}_{\nu}\) in slab \(i+1\).

\begin{equation}
\label{eq:velocity}
v_{\nu} = \frac{2ae}{\hbar} \mathrm{Im} \left( \chi^{i\dagger}_{\nu} \chi^{i+1}_{\nu} \right)
\end{equation}

The local current density of a particular mode between two grid points \((i,j)\) and \((i',j')\) is calculated using the current operator

\begin{equation}
\label{eq:current_operator}
 I_{\nu}((i,j) \rightarrow (i',j')) = 2 \mathrm{Im} \left( \chi^{i,j}_{\nu} \mathcal{H}_{i,j,i',j'} \chi^{i',j'}_{\nu} \right)
\end{equation}

\noindent \(\mathcal{H}_{i,j,i',j'}\) is the matrix element between grid points \((i,j)\) and \((i',j')\). The transmission of each mode at the Fermi level, \(T_{\nu} (E_F)\), is obtained directly from Kwant. We assume that the resistance is in the linear regime, since the Fermi window is narrow at low temperature and source drain bias is on the order of \(\mu\)V. This assumption essentially means that the edge states are in equilibrium with the bulk, since the linear regime resistance is an equilibrium property. The linear response resistance of the QPC at low temperature can be written as\cite{Datta1995}

\begin{equation}
\label{eq:qpc_resistance}
  R = \frac{h}{e^2} \left( \sum_{\nu} T_{\nu} \left( E_F \right) \right)^{-1}
\end{equation}

\section{Results} \label{sec:results}

\subsection{QPC resistance benchmark with experiment} \label{sec:benchmark}

\begin{figure}
  \includegraphics{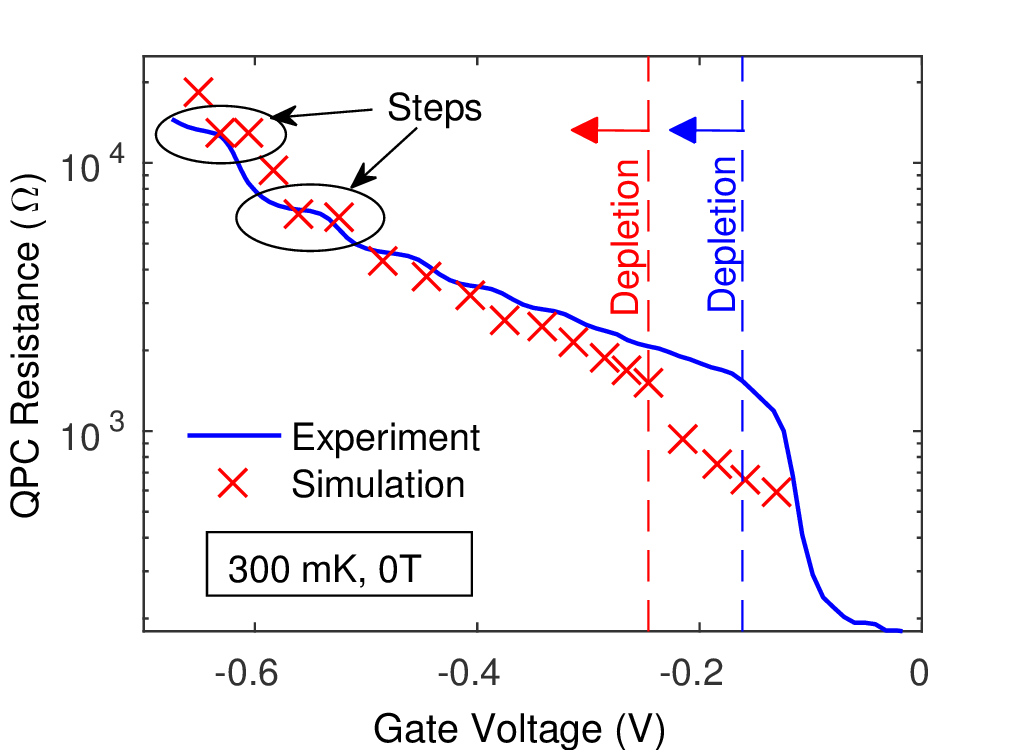}
  \caption{Comparison between experimentally measured and computed resistance of a \(300\) nm wide QPC. The measurement was done at \(300\) mK in \(0\) T magnetic field and using a constant AC current source of \(10\) nA. The red (light) and blue (dark) dashed lines indicate the depletion voltages in simulation and experiment respectively.}
  \label{fig:resistance_comparison_noB}
\end{figure}

Figure \ref{fig:resistance_comparison_noB} shows a comparison between measured and computed resistance for a \(300\) nm wide QPC for the \(91\) nm deep 2DEG heterostructure. The resistance measurement was done at \(300\) mK, with no magnetic field using a constant AC current source of \(10\) nA. The computed resistance shows a good match with the experimentally measured values. The discrepancies between calculated and measured resistance are near depletion (low negative voltage) and pinch-off (high negative voltage) regimes, and agreements are in between these regimes. The electron beam lithography system used to define the QPC has an effective resolution of approximately 20nm, so variations in the true width of the QPC on the order of 20nm are expected, which could lead to additional discrepancy.

The computed depletion voltage is about twice the measured value, because a simple one-level donor ionization model is used and physics of DX center formation is not captured. Also, below the depletion gate voltage the experimental resistance falls to zero, whereas in the simulations the boundary of the 2DEG is the simulation domain (which is much smaller than a Hall bar) leading to a minimum finite resistance. Due to this unphysical condition, simulated resistance isn't shown for gate voltages above the depletion point. Near the pinch-off, the source-drain bias is the highest and thus the accuracy of eq. \eqref{eq:qpc_resistance} is smaller than at lower voltages. The Thomas-Fermi approximation used to compute density of states laterally gives an inaccurate electrostatic potential in the QPC near pinch-off. Despite these minor discrepancies, the overall agreement between the experiment and simulation is satisfactory between the depletion and pinch-off regimes.

\subsection{Sub-band energy and sheet density profiles}

\begin{figure}
  \includegraphics{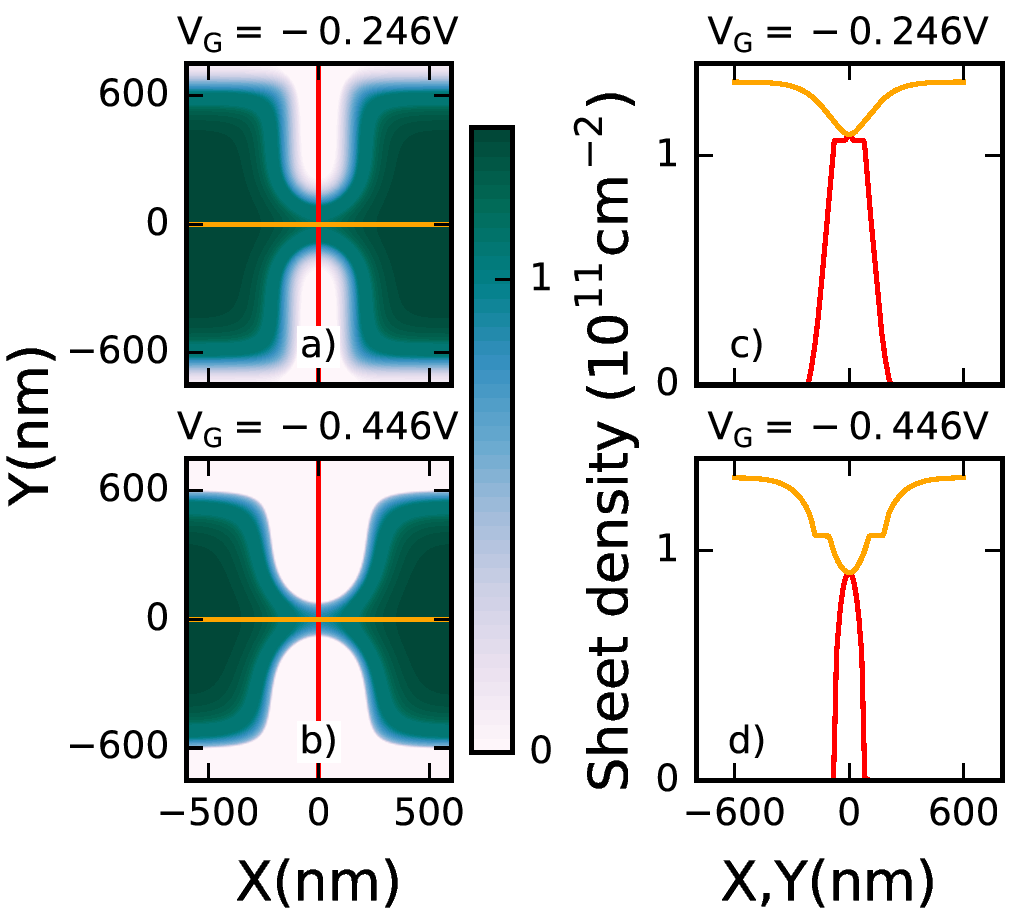}
  \caption{Sheet densities at different gate voltages for a \(300\) nm wide QPC, calculated using the Schr\"odinger-Poisson solver in a magnetic field of \(2.2\) T. Panels \textbf{c)} and \textbf{d)} show cuts along the X and Y axes, passing through the middle of the QPC. Sheet density in the bulk is \(1.34 \times 10^{11} \mathrm{cm}^{-2}\), and in the incompressible strip for \(n=1\) LL has a density of \(1.06 \times 10^{11} \mathrm{cm}^{-2}\). The incompressible strip can be seen as a light green region near the depleted 2DEG in \textbf{a)} and \textbf{b)}, and as flat region in \textbf{c)} and \textbf{d)}.}
  \label{fig:sheet_density_B}
\end{figure}

Figure \ref{fig:sheet_density_B} shows the sheet densities near the QPC for different gate voltages at a magnetic field of \(2.2\mathrm{T}\). For a sheet density of \(1.34 \times 10^{11} \mathrm{cm}^{-2}\), a single incompressible strip is expected in the presence of a magnetic field of \(2.2\mathrm{T}\). The incompressible strip can be seen as a band of light green with a density of around \(1\times 10^{11} \mathrm{cm}^{-2}\) near the edges. Using electrostatic simulations, the electron density in the middle of the QPC can be obtained for different gate voltages and magnetic fields. This helps in designing QPCs with the required channel width so that the velocity can be maximized keeping a certain filling fraction in the middle. The correctness of density and potential profiles can be verified by comparing conductance of the constriction for different gate voltages with experiments. The conductance can be calculated using quantum transport simulations.

\begin{figure}
  \includegraphics{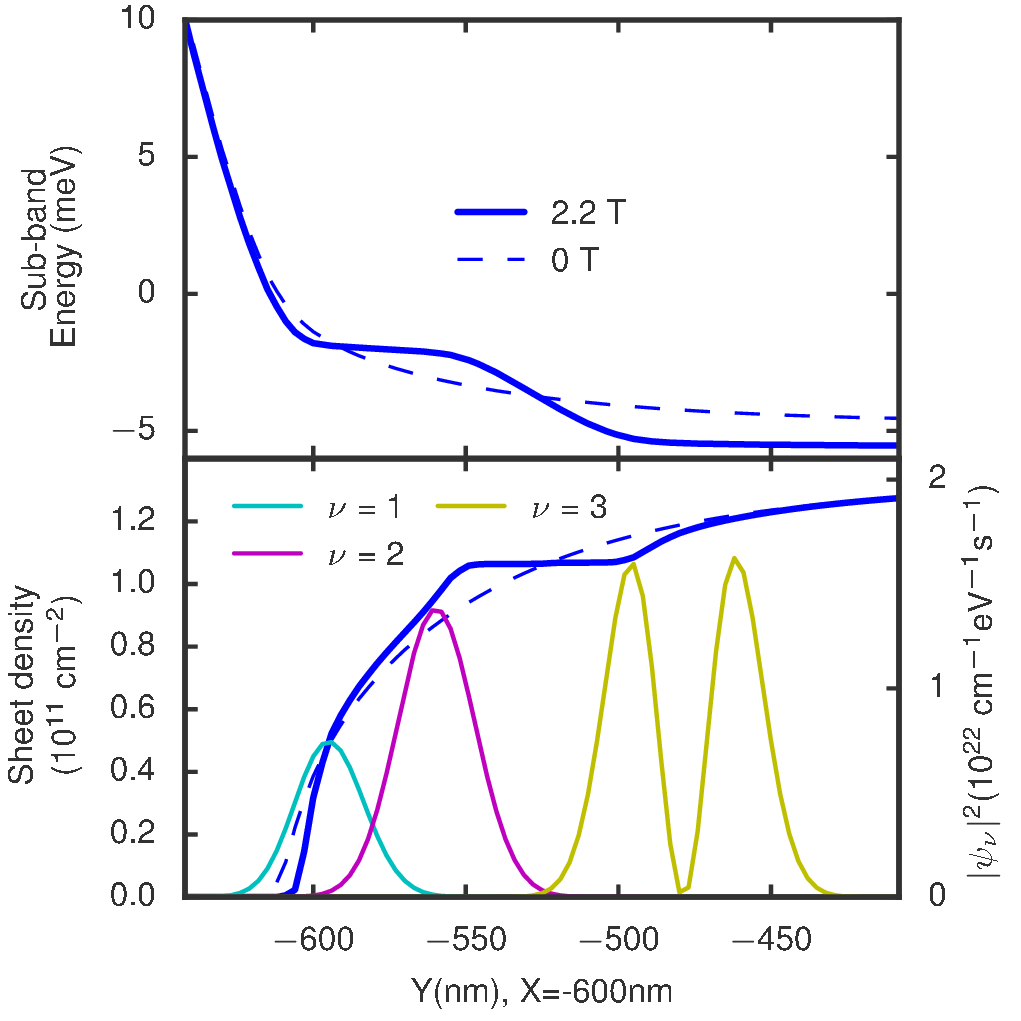}
  \caption{Sub-band energy (\(E_F = 0\)) and sheet density profiles near the edge of the 2DEG defined by depletion top gates at a magnetic field of 2.2 T (bulk filling factor \(\nu_{\mathrm{bulk}}=2.52\)) and gate voltage of -0.446 V compared to zero magnetic field values plotted using a dashed line. Edge state wavefunctions at the Fermi level for \(\nu=1\), 2 and 3 Landau levels obtained from quantum transport are also shown. }
  \label{fig:potential_density_traces_B}
\end{figure}

Figure \ref{fig:potential_density_traces_B} shows traces of the \(1^{st}\) sub-band energy and 2DEG sheet density obtained perpendicular to the 2DEG edge defined by the top gate for a magnetic field of 2.2 T (bulk filling factor \(\nu_{\mathrm{bulk}}=2.52\)) and a gate voltage of -0.446 V. The electron density in the presence of magnetic field forms the so called dipolar strips as predicted by Chklovskii et al.\cite{Chklovskii1992}. These dipolar strips form as a result of LLs crossing the Fermi level, and because of the dominating electrostatic forces due to the gates leading to a density profile that looks similar to the one without magnetic field. The region where electron density remains constant is the incompressible region, as the Fermi level lies between LLs where the density of states is zero. The incompressible regions in the sub-band energy plot have a finite potential drop across them. The regions where the sub-band energy is flat are called compressible regions because they are located where the LLs intersect the Fermi level. The electrostatics in these regions is similar to that of metals due to a large available density of states.

Figure \ref{fig:potential_density_traces_B} also shows the edge state wavefunctions for Landau levels \(\nu = 1\), 2 and 3 obtained from quantum transport simulations. \(\nu = 1\) corresponds to spin down and \(\nu = 2\) corresponds to spin up in the \(n = 1\) Landau level, therefore their wavefunctions have a single lobe. Here, \(n\)-Landau levels are spin-less and contain twice the density of states as the \(\nu\)-Landau levels. The \(\nu=3\) wavefunction lies in the \(n=2\) Landau level and has two lobes. These are the solutions for the Landau gauge. The plotted wavefunctions are normalized to carry a unit current (1 electron per second) such that their transmission equals 1. Therefore, the faster an edge state moves, the smaller its normalization. The wavefunctions clearly show that the edge states are present in the compressible regions.

\subsection{QPC conductance in the IQHE regime}

\begin{figure}
  \includegraphics{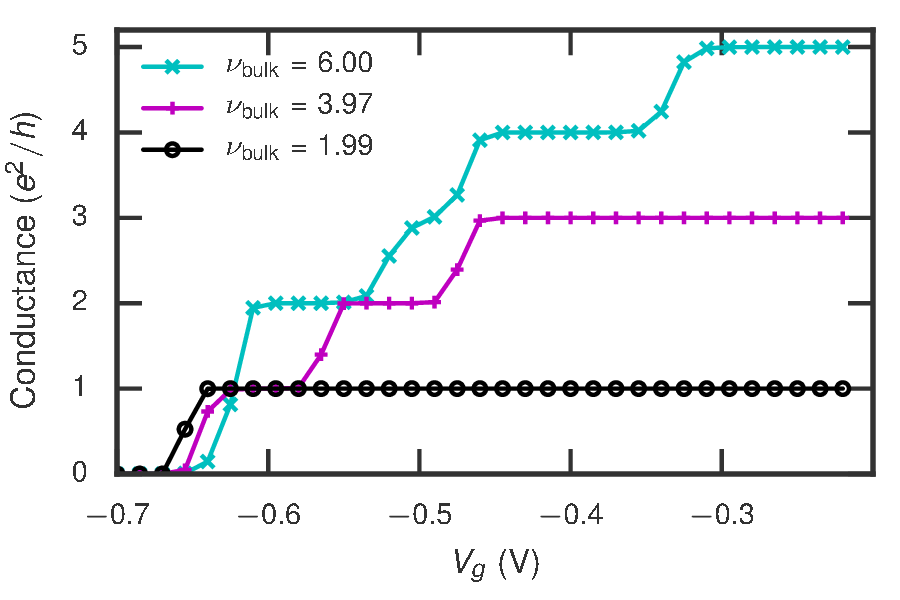}
  \caption{Conductance vs applied gate voltage of a 300 nm wide QPC at magnetic fields of 0.9T, 1.36T and 2.71T, and bulk filling factors of \(\nu_{\mathrm{bulk}} = 6.00\), \(\nu_{\mathrm{bulk}} = 3.97\) and \(\nu_{\mathrm{bulk}} = 1.99\) respectively. Each point represents an independent electrostatic simulation. The edge mode of even \(\nu^{th}\) Landau level starts conducting just above a bulk filling factor of \(\nu_{\mathrm{bulk}}=\nu\). This is why we have conductance corresponding to one less edge state than the filling factor.}
  \label{fig:conductance_comparison_B}
\end{figure}

The conductance is calculated by summing up the transmissions due to all the modes at the Fermi level. Figure \ref{fig:conductance_comparison_B} plots conductance of the QPC at three different magnetic fields.  The conductance shows exactly quantized plateaus in units of \(e^2/h\) when the modes are either completely transmitted or reflected by the QPC. We get \(\nu - 1\) conducting modes when the bulk filling factor is close to an even integer. This is because the modes in the \(\nu\) Landau level at that filling fraction have a very small velocity as we will show later in section \ref{sec:mag_field_dep_of_vel}. The electron density and filling factors are obtained from electrostatic simulations. The quantum transport simulations independently show that the extended states lie at the center of LL density of states, which means that the conducting edge states lie in the compressible regions.

\subsection{Edge state wavefunctions in the QPC scattering region}

\begin{figure*}
  \includegraphics[width=0.95 \textwidth ]{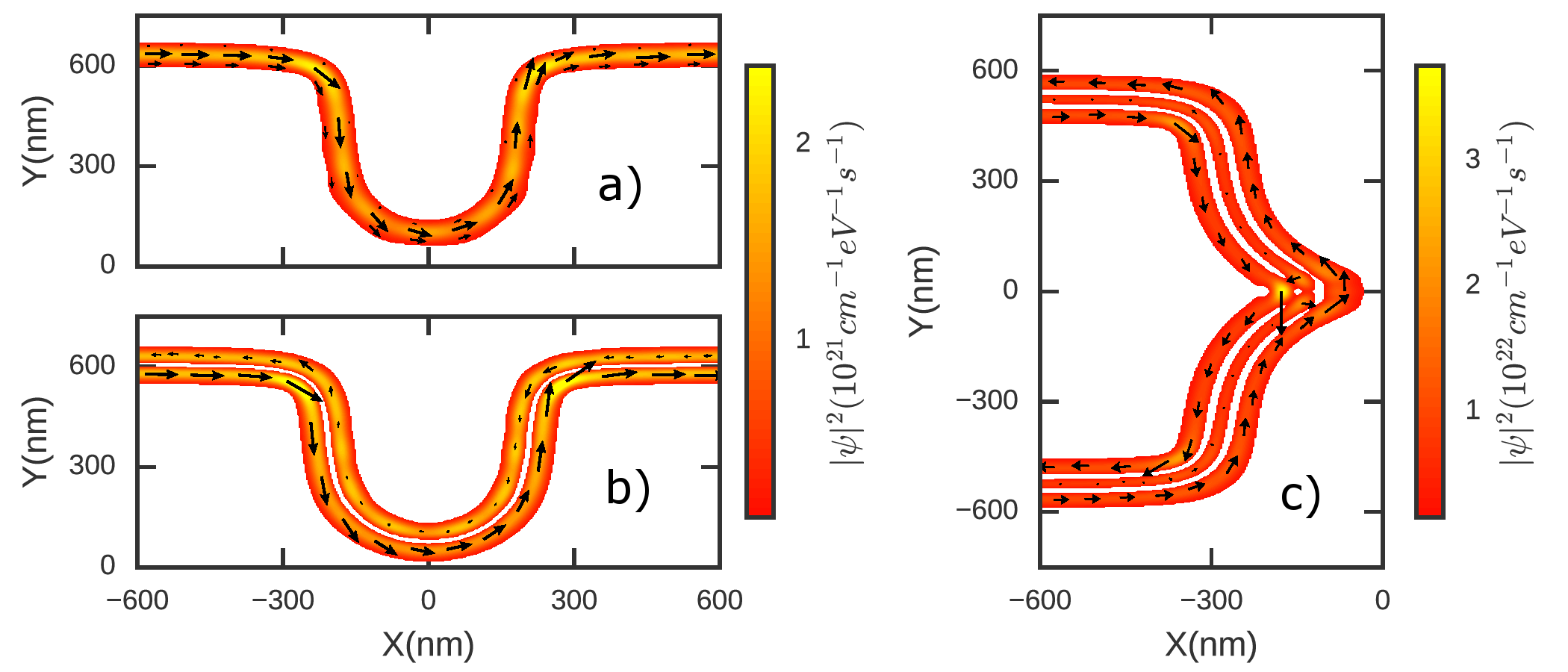}
  \caption{A plot of the 3 edge modes in the QPC structure in figure \ref{fig:structure_details} for \(n=1\) (\textbf{a}), \(n=2\) (\textbf{b}) and \(n=3\) (\textbf{c}) Landau levels at \(V_g=-0.342\mathrm{V}\) and \(B=0.9\mathrm{T}\). The arrows show the direction and relative magnitude of the current density. The conductance of QPC is \(2 \times 2e^2/h\), since the innermost mode is reflected by the QPC.}
  \label{fig:edge_modes_B}
\end{figure*}

Figure \ref{fig:edge_modes_B} shows edge mode wavefunctions at \(V_G=-0.342\)V and 0.9T (bulk filling factor of \(\nu_{\mathrm{bulk}} = 6\)) for a QPC. Figures \ref{fig:edge_modes_B}a) and b) have fully transmitted and \ref{fig:edge_modes_B}c) has fully reflected edge states respectively. In this simulation, the spin is neglected and we only calculate the wavefunctions for the spin-less Landau levels \(n\). The  wavefunctions for \(n=1\), 2 and 3 have 1, 2 and 3 lobes respectively, similar to the harmonic oscillator wavefunctions of the bulk 2D IQHE Hamiltonian in the Landau gauge. The wavefunctions for different spins in the same Landau level \(n\) have the same functional form. The current densities calculated using eq. \eqref{eq:current_operator} are shown using black arrows. Interestingly, the local current density changes direction for \(n=2\), 3 (figures \ref{fig:edge_modes_B} b), c)) and goes opposite to the direction of flow of current. Qualitatively, this may be understood semi-classically as the motion of the guiding center of the cyclotron orbit. Pile up of charge can be seen at the corners of the QPC defined 2DEG where the wavefunctions bend.


\section{Optimization of edge state velocity} \label{sec:optimization_of_velocity}

In this section we calculate velocities of edge states of different Landau levels as a function of gate voltage and magnetic field for different structures. We use equation \eqref{eq:velocity} for calculating the velocity of the edge states. We found that the velocity obtained from this equation is equal to \(\left< E \right> / B\), where \(\left< E \right>\) is the expectation value of the electric field for the edge state wavefunctions. The goal of this section is to design heterostructures and gates to obtain a strong electric field at the 2DEG edges that yield high edge state velocities. The velocity saturates even though electric field near the gate gets stronger, due to the finite width of the edge state wavefunction. Therefore, the correct metric for defining the strength of the electric field is the velocity of the edge states, or the expectation value of the electric field for the edge state wavefunctions \(\chi\). In this section the velocities of propagating modes are evaluated in the semi-infinite region, thus they are the injection velocities of propagating modes. Velocities calculated elsewhere in the device, but away from the middle region of the QPC, are found to be close to the injection velocities.

We consider four different structures as shown in figure \ref{fig:E_vs_n_structures} in an attempt to maximize the velocity of edge states. The first structure (fig. \ref{fig:E_vs_n_structures} a) is the same as fig. \ref{fig:structure_details} with top gates. In the second structure (fig. \ref{fig:E_vs_n_structures} b), the doping is moved closer to the 2DEG from 17-31 nm depth to 40-54 nm depth to increase the 2DEG bulk DOS. A depletion top gate is added on the bulk interferometer region to get the same bulk density as the first structure. The gate which defines the edge of the 2DEG is separated from the bulk depletion gate by 100 nm. The idea behind this design is the improve the electric field by making a higher 2DEG DOS available near the edge. The third structure (fig. \ref{fig:E_vs_n_structures} c) has the same heterostructure as the first structure, but uses trench gates with vertical trench walls which can be made using anisotropic etching techniques. Anisotropic etch trench gates which are etched past the doping region have a stronger effect on the 2DEG because the screening due to doping is removed. The fourth structure has double sided delta doping with trench gates etched past the quantum well but not the second doping layer. The bulk 2DEG density in this heterostructure is \(2.11 \times 10^{11}\) cm\textsuperscript{-2} as compared to \(1.34 \times 10^{11}\) cm\textsuperscript{-2} in the other three structures. Keeping the second doping layer helps pull the electrons closer to the gate and thus increases the edge electric field.

\subsection{Magnetic field dependence of velocity} \label{sec:mag_field_dep_of_vel}

\begin{figure*}
  \includegraphics[width=0.95 \textwidth ]{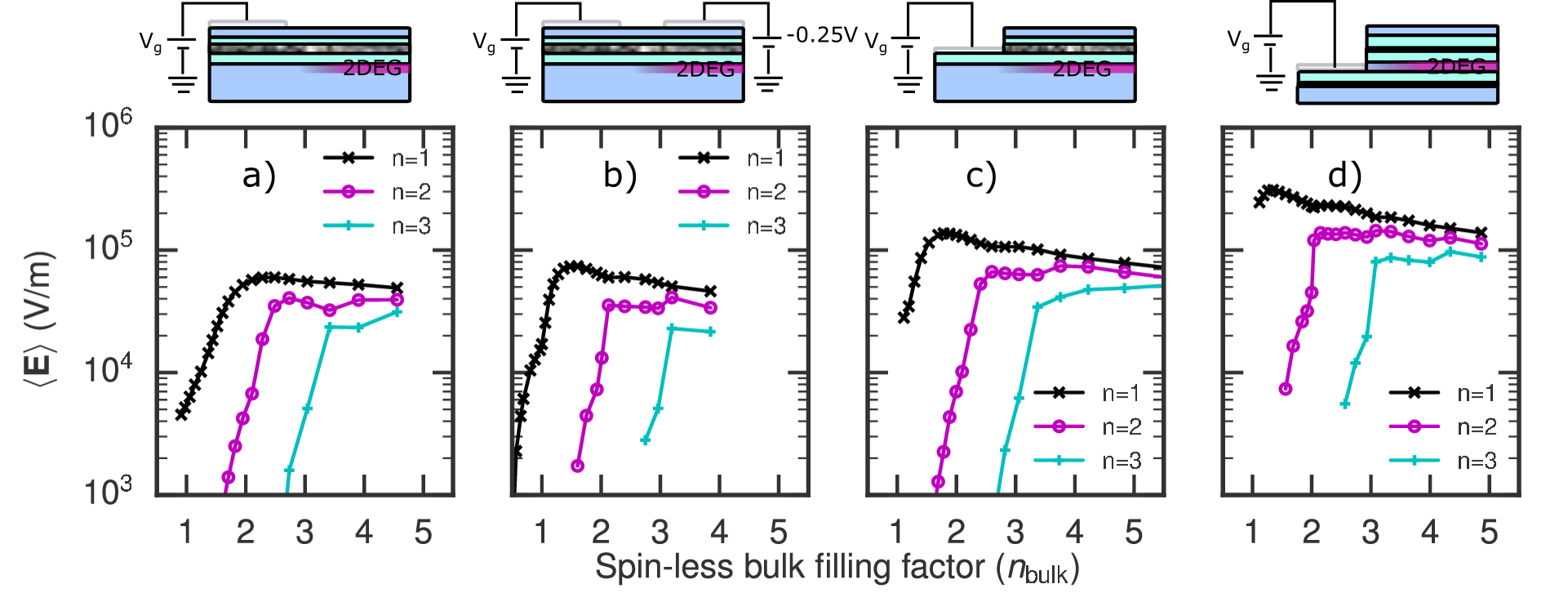}
  \caption{Electric field expectation values of edge modes of \(n=1,2,3\) LLs for four different structures are plotted as a function of the spin-less bulk filling factor. The structures are described in more detail in section \ref{sec:optimization_of_velocity}. Applied gate voltages (V\textsubscript{g}) are -0.34V, -0.54V, -0.1V and +0.1V for plots a), b), c) and d) respectively. }
  \label{fig:E_vs_n_structures}
\end{figure*}

\begin{figure}
  \includegraphics{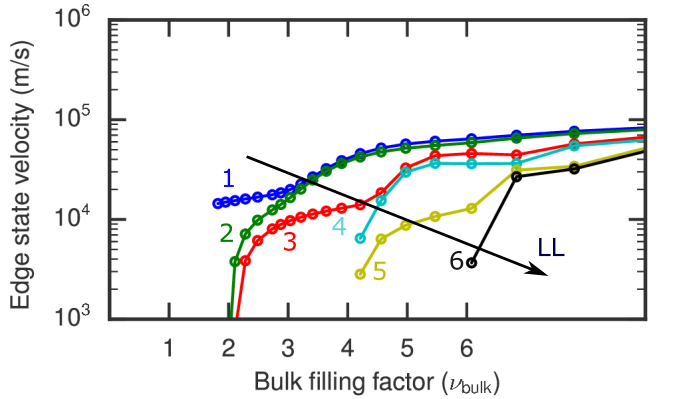}
  \caption{Velocity of edge states with spin for the structure in fig \ref{fig:E_vs_n_structures}a)  plotted as a function of the bulk filling factor \(\nu_{\mathrm{bulk}}\).}
  \label{fig:velocity_vs_nu_a}
\end{figure}

\begin{figure}
  \includegraphics{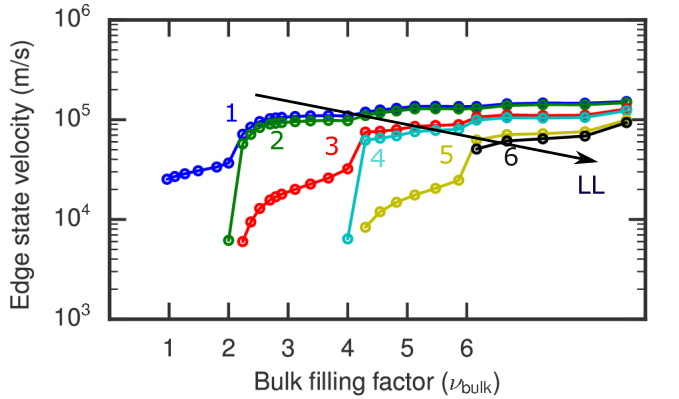}
  \caption{Velocity of edge states with spin for the structure in fig \ref{fig:E_vs_n_structures}d)  plotted as a function of the bulk filling factor \(\nu_{\mathrm{bulk}}\).}
  \label{fig:velocity_vs_nu_d}
\end{figure}

The expectation values of the electric field for edge states of the \(n\)-Landau levels are plotted as a function of the spin-less bulk filling factor for the four different structures in figure \ref{fig:E_vs_n_structures}. We plot the electric field expectation for the spin-less edge state wavefunctions in figure \ref{fig:E_vs_n_structures} to compare the compressible strips widths. The top gated structures require a negative bias on the top gate to deplete the 2DEG, whereas the 2DEG under trench gates is depleted at 0V because of the etched doping. A positive bias is applied in the fourth structure to pull the electrons towards the gate. It can be seen from the plots that the electric field expectation value goes to zero as the magnetic field is increased, for the edge states of Landau levels that are partially filled. The electric field expectation value for partially filled Landau levels starts dropping to zero close to half filled bulk Landau level, which means that the edge (extended) states lie at the center of the Landau level DOS.

The two features that define the sharpness of the edge are the maximum value of \(\left< \mathbf{E} \right> \) and its slope as a function of the filling factor. Both these features depend on the width of compressible region. The compressible region is narrower in a structure as compared to another at a particular magnetic field (equal sub-band energy drop across the compressible regions in the two structures), when \(\left< \mathbf{E} \right> \) is stronger in that structure. Due to this, the slope of \(\left< \mathbf{E} \right> \) as a function of the bulk filling factor is also steeper because the width of the compressible region decreases to a smaller value. The electric field in trench gated double delta doped structure is the highest among the four structures (figure \ref{fig:E_vs_n_structures}).

\begin{figure*}
  \includegraphics[width=0.95 \textwidth ]{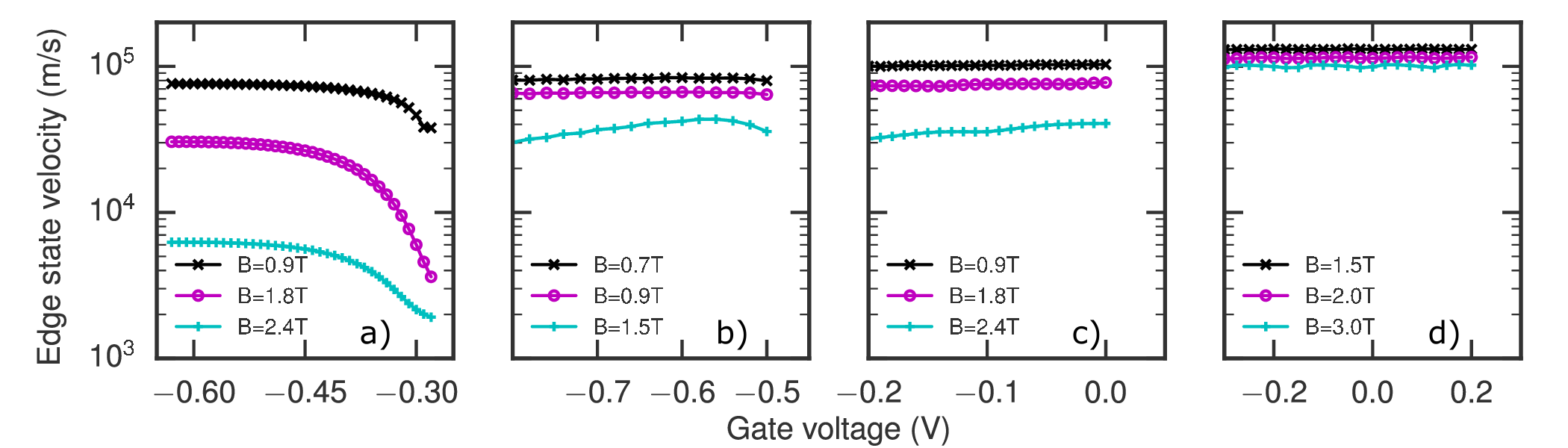}
  \caption{Velocities of edge modes for outermost (\(n=1\)) \(n-\)Landau level in same four structures as figure \ref{fig:E_vs_n_structures} at different magnetic fields are plotted as a function of the gate voltage. The spin-less bulk filling factors for the plotted magnetic fields are: a) \(n_{\mathrm{bulk}} = \) 3.04, 1.51 and 1.14; b) \(n_{\mathrm{bulk}} = \) 2.96, 2.12 and 1.27; c) \(n_{\mathrm{bulk}} = \) 3.75, 1.88 and 1.40; d) \(n_{\mathrm{bulk}} = \) 2.93, 2.14 and 1.44; respectively. }
  \label{fig:v_vs_Vg_structures}
\end{figure*}

Figures \ref{fig:velocity_vs_nu_a} and \ref{fig:velocity_vs_nu_d} plot the edge state velocities for structures \ref{fig:E_vs_n_structures}a) and \ref{fig:E_vs_n_structures}d) respectively when the spin is included in the Hamiltonian. With spin included the edge state wavefunctions, which were in the center of the compressible region when spin was excluded, split into two and move towards opposite ends of the compressible region. This can be seen in figure \ref{fig:potential_density_traces_B} for \(\nu=1, 2\). This is due to the Zeeman energy and the large effective spin splitting due to interactions\cite{Huang2013}. Edge states with opposite spin from consecutive \(n\)-Landau levels lie close to each other, e.g. in fig. \ref{fig:potential_density_traces_B} for \(\nu=2, 3\). Therefore the velocity of even \(\nu\)-Landau level edge states goes to zero just below the corresponding \(\nu\) bulk filling factor as the wavefunction is close to the inner edge of the compressible region. The velocity of odd \(\nu\)-Landau level edge states goes to zero close to \(\nu-1\) bulk filling factor as the wavefunction is close to the outer edge of the compressible region. This explains the maximum conductance in fig. \ref{fig:conductance_comparison_B} (conductance plot for the QPC in the IQHE regime). From these plots we can infer that the edge state velocity has an upper limit due to a finite width of the edge state wavefunctions. The velocities are also affected by the electrostatics. This can be clearly seen for the inner edge states, whose velocity decreases close to the even integer \(\nu_{\mathrm{bulk}}\) bulk filling factor and forms plateaus in between.

Our results agree qualitatively and show a reasonable quantitative agreement with observations of McClure et al.\cite{McClure2009}, as well as with the experimental observations made by Gurman et al.\cite{Gurman2016} at high filling (\(\nu_{\mathrm{bulk}} > 3\)). The velocities predicted by our model fall reasonably close to the experimentally measured values; for example \(\sim 0.5-1.5 \times 10^5\) m/s measured by Mcclure et al\cite{McClure2009} and \(\sim 2-8 \times 10^4\) m/s measured by Gurman et al.\cite{Gurman2016}. We predict that the velocities in the trench gated structures are higher only by a factor between 1-2 and as much as 10 at certain filling factors for the outer edge. Additionally, our results predict that the velocity of the inner spin-split \(\nu = 2\) Landau level drops to zero below the bulk filling factor \(\nu_{\mathrm{bulk}}=2\), which is consistent with the findings by Gurman et al.\cite{Gurman2016} that the visibility of interference of this inner Landau level vanishes below a bulk filling factor of \(\nu_{\mathrm{bulk}}=2\). However, our model does not predict the experimental measurements that can be linked with e-e interactions and inter-edge scattering, such as the non-monotonic dependence of the velocity and interference visibility observed by Gurman et al. at low bulk filling factors (\(\nu_{\mathrm{bulk}} < 2.5\)).

\subsection{Gate voltage dependence of velocity}

The edges of the 2DEG in the interferometer region are defined by negatively biased top gates, or trench gates. The 2DEG in the top gated structures is not depleted till a certain negative gate voltage is reached as the doping layer screens the top gate. Trench gated structures on the other hand have depleted 2DEG irrespective of the applied gate voltage when the doping layer is also etched. A positive gate bias can then be applied on the trench gates to pull the edge of the 2DEG closer to the lithographically defined edge to possibly get a larger edge electric field. In this sub-section we compare the edge state velocities for top and trench gated structures and study their dependence on the gate voltage.

Figure \ref{fig:v_vs_Vg_structures} plots the velocity of the outermost \(n\)-Landau level edge mode at different magnetic fields as a function of the gate voltage for the four structures studied in the previous sub-section. The velocity for the top gated structure (fig. \ref{fig:v_vs_Vg_structures} a)) increases as more negative voltage is applied on the gate and saturates at a certain value. This shows that the 2DEG is screened by the doping layer and the edge potential is the steepest when the doping layer is depleted. Fig. \ref{fig:v_vs_Vg_structures} b) shows velocities at high negative gate voltages in the saturated regime. We can still see a peak in the velocity near -0.6V for B=1.5T. This is because the edge state moves from under the outer top gate to under the inner top gate while passing through the region with exposed surface. This shows that the region with exposed surface has a larger electric field due to a higher availability of local density of states. This effect is not seen at B=0.7T and B=0.9T because the wavefunctions get wider as magnetic field is reduced and the electric field is averaged out over a larger area. Figs. \ref{fig:v_vs_Vg_structures} c) and d) are for the trench gated structures. We propose that the velocity increases in c) the as the gate voltage becomes more positive for B=2.4T, because the edge state wavefunction is pulled closer to the sharper edge potential near the lithographic edge. This effect is less pronounced at lower magnetic fields where the wavefunctions are wider. d) shows a very small to no dependence on the gate voltage. Our understanding is that this is due to the positively charged lower doping layer pulling the edge state wavefunction closer to the lithographic edge. The gate voltage only changes the location of the edge state and not the electric field expectation value.

\section{Summary}

We have presented a simulation method for comparing the edge state velocity in different structures and designing the heterostructures for increasing the velocity. We have implemented a way of solving Schr\"odinger and Poisson equations self-consistently in QPCs defined on GaAs/AlGaAs heterostructures, for obtaining the electron density and electrostatic potential in the IQHE. A set of 1-dimensional wavefunctions are solved for the interfacial quantum well on a lateral 2-dimensional grid, and the Thomas-Fermi Approximation (TFA) is used to calculate lateral density of states to get the 3-dimensional electron density used in the electrostatic simulations. The broadening of Landau levels due to disorder and various other effects is considered in the TFA using a Gaussian broadening of LLs. Electrostatic simulations show the formation of compressible and incompressible regions near the edge of the 2DEG. The sub-band energies of the quantum well obtained from self-consistent simulations are used in 2-dimensional quantum transport calculations to get the transmission, edge state wavefunctions, velocities and current densities.

We show realistic sheet density, sub-band energy and edge state wavefunction profiles for QPCs obtained from the model. We benchmark our model by comparing the calculated resistance with the measured resistance for the same structure fabricated experimentally. We obtain the edge state wavefunctions from the quantum transport simulations, which represent the solutions of the 2-dimensional IQHE Hamiltonian in the Landau gauge and for a spatially varying electric field.

The velocity of edge states calculated using quantum transport simulations matches with \(\left< \mathbf{E} \right> / B\), where \(\left< \mathbf{E} \right>\) is the electric field expectation value for the edge states. The edge state velocity has an upper limit due to a finite width of the edge state wavefunctions. We also compare the magnetic field and gate voltage dependence of the edge state velocity for different structures. We conclude that the velocity in the double delta doped anistropic etched trench gated structures is the highest among the four structures considered.



Our results can be used to understand some of the visibility and velocity measurements of electronic interferometers operating in the integer quantum Hall regime\cite{Gurman2016,McClure2009}. The device designs we have proposed may lead to improved edge state velocity and thus improved performance of future interferometers, and may enable the observation of interference in the fractional quantum Hall regime. 

\begin{acknowledgments}
This research has been funded by Purdue Center for Topological Materials. This research is part of the Blue Waters sustained-petascale computing project, which is supported by the National Science Foundation (awards OCI-0725070 and ACI-1238993) and the state of Illinois. Blue Waters is a joint effort of the University of Illinois at Urbana-Champaign and its National Center for Supercomputing Applications. H. Sahasrabudhe thanks Prof. Supriyo Datta for helpful discussions on boundary conditions for the Schr\"odinger equation.
\end{acknowledgments}


\bibliography{Mendeley}

\end{document}